# Quantum Query Complexity of Multilinear Identity Testing

V. Arvind and Partha Mukhopadhyay

Institute of Mathematical Sciences

C.I.T Campus,Chennai 600 113, India

{`arvind,partham`}`@imsc.res.in`

### Abstract

Motivated by the quantum algorithm in [MN05] for testing commutativity of black-box groups, we study the following problem: Given a black-box finite ring $R = \langle r_1, \cdots, r_k \rangle$ where $\{r_1, r_2, \cdots, r_k\}$ is an additive generating set for $R$ and a multilinear polynomial $f(x_1, \cdots, x_m)$ over $R$ also accessed as a black-box function $f : R^m \to R$ (where we allow the indeterminates $x_1, \cdots, x_m$ to be commuting or noncommuting), we study the problem of testing if $f$ is an *identity* for the ring $R$. More precisely, the problem is to test if $f(a_1, a_2, \cdots, a_m) = 0$ for all $a_i \in R$.

- We give a quantum algorithm with query complexity $O(m(1 + \alpha)^{m/2} k^{\frac{m}{m+1}})$ assuming $k \geq (1 + 1/\alpha)^{m+1}$. Towards a lower bound, we also discuss a reduction from a version of $m$-collision to this problem.

- We also observe a randomized test with query complexity $4^m mk$ and constant success probability and a deterministic test with $k^m$ query complexity.

## 1 Introduction

For any finite ring $(R, +, \cdot)$ the ring $R[x_1, x_2, \cdots, x_m]$ is the ring of polynomials in commuting variables $x_1, x_2, \cdots, x_m$ and coefficients in $R$. The ring $R\{x_1, x_2, \cdots, x_m\}$ is the ring of polynomials where the indeterminates $x_i$ are *noncommuting*. By noncommuting variables, we mean $x_i x_j - x_j x_i \neq 0$ for $i \neq j$.

For the algorithmic problem we study in this paper, we assume that that the elements of the ring $(R, +, \cdot)$ are uniformly encoded by binary strings of length $n$ and $R = \langle r_1, r_2, \cdots, r_k \rangle$ is given by an additive generating set $\{r_1, r_2, \cdots, r_k\}$. That is,

$$R = \{\sum_i \alpha_i r_i \mid \alpha_i \in \mathbb{Z}\}.$$

Also, the ring operations of $R$ are performed by black-box oracles for addition and multiplication that take as input two strings encoding ring elements and output their sum or product (as the case may be). Additionally, we assume that the zero element of $R$ is encoded by some fixed string. We now define the problem which we study in this paper.

**The Multilinear Identity Testing Problem** (MIT): The input to the problem is a black-box ring $R = \langle r_1, \cdots, r_k \rangle$ given by an additive generating set, and a multilinear polynomial $f(x_1, \cdots, x_m)$ (in the ring $R[x_1, \cdots, x_m]$ or the ring $R\{x_1, \cdots, x_m\}$) that is also given by black-box access. The problem is to test if $f$ is an *identity* for the ring $R$. More precisely, the problem is to test if $f(a_1, a_2, \cdots, a_m) = 0$ for all $a_i \in R$.

A natural example of an instance of this problem is the bivariate polynomial $f(x_1, x_2) = x_1 x_2 - x_2 x_1$ over the ring $R\{x_1, x_2\}$. This is an identity for $R$ precisely when $R$ is a commutative ring. Clearly, it suffices to check if the generators commute with each other which gives a naive algorithm that makes $O(k^2)$ queries to the ring oracles.

Given a polynomial $f(x_1, \cdots, x_m)$ and a black-box ring $R$ by generators, we briefly recall some facts about the complexity of checking if $f = 0$ is an identity for $R$. The problem can be NP-hard when the number of indeterminates $m$ is unbounded, even when $R$ is a fixed ring. To see this, notice that a 3-CNF formula $F(x_1, \cdots, x_n)$ can be expressed as a $O(n)$ degree multilinear polynomial $f(x_1, x_2, \cdots, x_n)$ over $\mathbb{F}_2$, by writing $F$ in terms of addition and multiplication over $\mathbb{F}_2$. It follows that $f = 0$ is an identity for $\mathbb{F}_2$ if and only if $F$ is an unsatisfiable formula.

We remark that a closely related problem is Polynomial Identity Testing (PIT). For PIT we ask whether the polynomial $f(x_1, \cdots, x_m)$ is the zero polynomial, which is a stronger property. To see the difference, consider a standard example: For a prime $p$, notice that $x^p - x = 0$ is an identity for $\mathbb{F}_p$ but $x^p - x$ is not a zero polynomial in $\mathbb{F}_p[x]$. However, when the ring $R$ is a *field* $\mathbb{F}$ and the degree of $f$ is smaller than the size of the field $\mathbb{F}$ then the two problems coincide as a consequence of the Schwartz-Zippel lemma [Sch80, Zip79]. More precisely, $f = 0$ is an identity for $\mathbb{F}$ if and only if $f$ is the zero polynomial.

When $f$ is given by an *arithmetic circuit* then PIT is known to be in randomized polynomial time over fields [Sch80, Zip79] and even finite commutative rings with unity [AB03, AMS08]. This is quite unlike MIT which can be NP-hard for polynomials over small fields as already observed above.

On the other hand, when $f$ is given by black-box access as a function $f : R^m \to R$ then there is no way to distinguish between the problems PIT and MIT. Algorithmically, they coincide.

Over the years, Polynomial Identity Testing has emerged as an important algorithmic problem [AB03, KI03]. Due to its significance in complexity theory, PIT has been actively studied in recent years [DS06, KS07, RS05].

In this paper we focus on the *query complexity* of *multilinear identity testing* (MIT). In our query model, each ring operation, which is performed by a query to the ring oracle, is of unit cost. Furthermore, we consider each evaluation of $f(a_1, \cdots, a_m)$ to be of unit cost for a given input $(a_1, \cdots, a_m) \in R^m$. This model is reasonable because we consider $m$ as a parameter that is much smaller than $k$.

Our goal is to find upper and lower bounds on the query complexity for the problem. We are interested in the query complexity for both classical and quantum computation. The main motivation for our study is a result of Magniez and Nayak in [MN05], where the authors study the quantum query complexity of group commutativity testing: Let $G$ be a finite black-box group given by a generating set $g_1, g_2, \cdots, g_k$ and group operations are performed by a group oracle. The algorithmic task is to check if $G$ is commutative. For this problem the authors in [MN05] give a quantum algorithm with query complexity $O(k^{2/3} \log k)$ and time complexity $O(k^{2/3} \log^2 k)$. Furthermore a $\Omega(k^{2/3})$ lower bound for the quantum query complexity is also shown. The main technical tool for their upper bound result was a method of quantization of random walks first showed by Szegedy [Sze04]. More recently, Magniez et al in [MNRS07] discovered a simpler and improved description of Szegedy's method.

Our starting point is the observation that the Magniez-Nayak result [MN05] for group commutativity can also be easily seen as a commutativity test for arbitrary finite black-box rings. If $R = \langle r_1, \cdots, r_k \rangle$ is a finite black-box ring and $f$ is the bivariate polynomial $f(x_1, x_2) = x_1 x_2 - x_2 x_1$ over the polynomial ring $R\{x_1, x_2\}$ ($x_1, x_2$ do not commute). Testing if $f = 0$ is an identity for $R$ is testing if $R$ is commutative. It turns out that the Magniez-Nayak results can be easily adapted to obtain similar upper and lower bounds for the quantum query complexity of the problem. Motivated by this connection we study the problem of testing multilinear identities for any black-box ring. We crucially need the multilinearity condition to generalize a result of Pak [Pak00] to multilinear polynomials. Given a black-box group $G = \langle g_1, \cdots, g_k \rangle$ by a generating set, Pak shows in [Pak00] that it suffices to plug in random subproducts of the generators for variables $g$ and $h$ in the equation $gh = hg$ to check for commutativity. Pak shows that for such random subproducts $gh \neq hg$ with constant probability if $G$ is nonabelian. If $R = \langle r_1, \cdots, r_k \rangle$ is a finite black-box ring given by an additive generating set, Pak's result can be easily modified to show the following: if we plug in *random subsums* of the generators $r_1, \cdots, r_k$ for the variables $x_1$ and $x_2$ in the polynomial $x_1 x_2 - x_2 x_1$, then for noncommutative rings $R$ we will have $x_1 x_2 - x_2 x_1 \neq 0$ with constant probability. We prove a generalization of this property for any multilinear polynomial $f(x_1, \cdots, x_m)$. Then, using the Magniez-Nayak technique adapted suitably, we show a quantum algorithm for this problem with quantum query complexity $O(m(1 + \alpha)^{m/2} k^{\frac{m}{m+1}})$ when $(1 + 1/\alpha)^{m+1} \leq k$.

For the lower bound result Magniez and Nayak show a reduction from UNIQUE COLLISION: let $f$ be a function from $\{1, 2, \cdots, k\}$ to $\{1, 2, \cdots, k\}$ given as a oracle, with the promise is that either there exists a unique collision pair $x \neq y$ such that $f(x) = f(y)$ or $f$ is a permutation. It is known from earlier work [AS04, Kut05, Amb05] that the quantum query complexity of UNIQUE COLLISION is $\Omega(k^{2/3})$. In fact Magniez and Nayak define a variant of UNIQUE COLLISION problem, which they call UNIQUE SPLIT COLLISION problem: Assume $k$ is even. Then, in the Yes instances, one element of the colliding pair has to come from $\{1, \cdots, k/2\}$ and the other from $\{k/2 + 1, \cdots, k\}$. Then their paper shows a reduction from UNIQUE COLLISION to UNIQUE SPLIT COLLISION and finally a reduction from UNIQUE SPLIT COLLISION to group commutativity testing.

We show a reduction to a somewhat more general version of MIT from a problem that is closely related to the m-COLLISION problem studied in quantum computation. Given a function $f : \{1, 2, \cdots, k\} \rightarrow \{1, 2, \cdots, k\}$ as an oracle and a positive integer $m$, the task is to determine if there is some element in the range of $f$ with exactly $m$ pre-images. More precisely, is there an $i \in [k]$ such that $|f^{-1}(i)| = m$? We define a new problem closely related to m-COLLISION problem, that we call m-SPLIT COLLISION problem. Here we divide the numbers $1, 2, \cdots, k$ into $m$ consecutive equal-sized intervals (assume $k$ is a multiple of $m$) and ask if there is some element in the range of $f$ with exactly one pre image in each of the $m$ intervals. We show a reduction from m-SPLIT COLLISION to a general version of MIT. We do not know an explicit lower bound for the quantum query complexity of m-SPLIT COLLISION (unlike UNIQUE SPLIT COLLISION in [MN05]). The reduction of UNIQUE SPLIT COLLISION to group commutativity testing problem in [MN05] directly gives a $\Omega(k^{2/3})$ lower bound for the quantum query complexity of the general version of MIT. However, we do not have a stronger lower bound. Ideally, we would like to have a dependence of $m$ in the exponent of $k$.

Our reduction from m-SPLIT COLLISION to MIT uses ideas from automata theory to construct a suitable black-box ring. Recently, in [AMS08] we used similar ideas to give a new deterministic polynomial time identity testing (PIT) algorithm for arithmetic circuits computing sparse and small degree multivariate polynomial over noncommuting variables.

**Remark.** Ambainis in [Amb04] showed the quantum upper bound of $O(k^{m/m+1})$ for the m-COLLISION problem. But $\Omega(k^{2/3})$ is the best known quantum lower bound for m-COLLISION for $m = 2$ [AS04]. The quantum query complexity of m-COLLISION has been open for some years.

There is a randomized reduction from m-COLLISION to m-SPLIT COLLISION with success probability $e^{-m}$: let $f : [k] \rightarrow [k]$ be a 'yes' instance of m-COLLISION, and suppose $f^{-1}(i) = \{i_1, i_2, \cdots, i_m\}$. To reduce this instance to m-SPLIT COLLISION we pick a random $m$-partition $I_1, I_2, \cdots, I_m$ of the domain $[k]$ with each $|I_j| = k/m$. Clearly, with probability $e^{-m}$ the set $\{i_1, i_2, \cdots, i_m\}$ will be a split collision for the function $f$. Consequently, showing a quantum lower bound of $\Omega(k^\alpha)$ for m-COLLISION will imply a quantum lower bound of $\Omega(k^\alpha/e^m)$ for m-SPLIT COLLISION and hence to MIT.

## 2 Black-box rings and the quantum query model

As already explained, the ring operations (addition and multiplication) for a black-box ring are performed by querying a ring oracle. We can modify the definition of black-box ring operations by making them unitary transforms that can be used in quantum algorithms. For a black-box ring $R$, we have two oracles $O_R^a$ and $O_R^m$ for addition and multiplication respectively. For any two ring elements $r, s$, and a binary string $t \in \{0, 1\}^n$ we have $O_R^a|r\rangle|s\rangle = |r\rangle|r + s\rangle$ and $O_R^m|r\rangle|s\rangle|t\rangle = |r\rangle|s\rangle|rs \oplus t\rangle$, where the elements of $R$ are encoded as strings in $\{0, 1\}^n$. Notice that $O_R^a$ is a reversible function by virtue of $(R, +)$ being an additive group. On the other hand, $(R, \cdot)$ does not have a group structure. Thus we have made $O_R^m$ reversible by defining it as a 3-place function $O_R^m : \{0, 1\}^{3n} \to \{0, 1\}^{3n}$. When $r$ or $s$ do not encode ring elements these oracles can compute any arbitrary string.

The query model in quantum computation is a natural extension of classical query model. The basic difference is that a classical algorithm queries deterministically or randomly selected basis states, whereas a quantum algorithm can query a quantum state which is a suitably prepared superposition of basis states. For a black-box ring operation the query operators are simply $O_R^a$ and $O_R^m$ (as defined above). For an arbitrary oracle function $F : X \to Y$, the corresponding unitary operator is $O_F : |g\rangle|h\rangle \to |g\rangle|h \oplus F(g)\rangle$. In the query complexity model, we charge unit cost for a single query to the oracle and all other computations are free. We will assume that the input black-box polynomial $f : R^m \to R$ is given by such an unitary operator $U_f$.

All the quantum registers used during the computation can be initialised to $|0\rangle$. Then a $k$-query algorithm for a black-box ring is a sequence of $k + 1$ unitary operators and $k$ ring oracle operators: $U_0, Q_1, U_1, \cdots, U_{k-1}, Q_k, U_k$ where $Q_i \in \{O_R^a, O_R^m, O_F\}$ are the oracle queries and $U_i$'s are unitary operators. The final step of the algorithm is to measure designated qubits and decide according to the measurement output.

## 3 Quantum Algorithm for multilinear Identity Testing

In this section we describe our quantum algorithm for multilinear identity testing MIT. Our algorithm is motivated by (and based on) the group commutativity testing algorithm of Magniez and Nayak [MN05]. We briefly explain the algorithm of Magniez and Nayak. Their problem was the following: given a black-box group $G$ by a set of generators $g_1, g_2, \cdots, g_k$, the task is to find nontrivial upper bound on the quantum query complexity to determine whether $G$ is commutative. The group operators (corresponding to the oracle) are $O_G$ and $O_{G^{-1}}$.

Note that for this problem, there is a trivial classical algorithm (so as quantum) of query complexity $O(k^2)$. In an interesting paper Pak showed a classical randomized algorithm of query complexity $O(k)$ for the same problem [Pak00]. Pak's algorithm is based on the following observation (Lemma 1.3 in [Pak00]): consider a subproduct $h = g_1^{e_1} g_2^{e_2} \cdots g_k^{e_k}$ where $e_i$'s are picked uniformly at random from $\{0, 1\}$. Then for any proper subgroup $H$ of $G$, $\text{Prob}[h \notin H] \geq 1/2$.

One important step of the algorithm in [MN05] is a generalization of Pak's lemma. Let $S_\ell$ be the set of all distinct element $\ell$ tuples of elements from $\{1, 2, \cdots, k\}$. For $u = (u_1, \cdots, u_\ell)$, define $g_u = g_{u_1} \cdot g_{u_2} \cdots g_{u_\ell}$. Let $p = \frac{\ell(\ell-1)+(k-\ell)(k-\ell-1)}{k(k-1)}$.

**Lemma 3.1** *[MN05] For any proper subgroup $K$ of $G$, $\text{Prob}_{u \in S_\ell}[g_u \notin K] \geq \frac{1-p}{2}$*

As a simple corollary of this lemma, Magniez and Nayak show in [MN05] that if $G$ is nonabelian then for randomly picked $u$ and $v$ from $S_\ell$ the elements $g_u$ and $g_v$ will not commute with probability at least $\frac{(1-p)^2}{4}$.

Thus, for noncommutative $G$ there will be at least $\frac{(1-p)^2}{4}$ fraction of noncommuting pairs $(u, v)$. Call such pairs as "marked pairs". Next, their idea is to do a random walk in the space of all pairs and hit a marked pair quickly (i.e. using only a few queries to the group oracle). They achieved this by defining a random walk and quantizing it using [Sze04, MNRS07]. The random walk consists of two independent random walks on $S_\ell$. For each $u \in S_\ell$, they maintain a binary tree $t_u$ whose leaves corresponds to $g_{u_1}, g_{u_2}, \cdots, g_{u_\ell}$ and the internal nodes corresponds to the group product of its two children. So $g_u$ is computed at the root of $t_u$. The description of the random walk is simple. Suppose the state is $u \in S_\ell$ at some stage. With probability $1/2$ the walk will stay at $u$ (this ensures the ergodicity of the walk) and with probability $1/2$ do the following: Pick $i$ uniformly at random from $1, 2, \cdots, \ell$ and pick $j$ uniformly at random from $1, 2, \cdots, k$. If $j$ is already equal to some $u_m$, exchange $u_i$ and $u_m$. Otherwise set $u_i = j$. Recompute the group operations at the nodes of $t_u$ which are affected by this substitution. It is easy to see that $t_u$ can be updated using only $O(\log \ell)$ queries to group oracle. Using a coupling argument of Markov Chain it is shown in [MN05] that the spectral gap $\delta$ of this random walk is at least $\frac{1}{8e\ell \log \ell}$. Since the random walk is ergodic its stationary distribution will be uniform. So the fraction of the marked states (pairs) in $S_\ell \times S_\ell$ will be at least $\frac{(1-p)^2}{4}$. Now they invoke Szegedy's result to perform a quantum walk on $S_\ell \times S_\ell$ and hit a marked element pair. We recall the statement of Szegedy's theorem. (For a detailed explanation see the section 2.3 of [MN05])

**Theorem 3.2** *[Sze04] Let $P$ be the transition matrix of a Markov Chain on a graph $G = (V, E)$ and $\delta$ be the spectral gap of $P$. Also let $M$ be the set of all marked vertices in $V$ and $|M|/|V| \geq \epsilon > 0$, whenever $M$ is nonempty. Then there is a quantum algorithm which determines whether $M$ is nonempty with constant success probability and query complexity $S + O((U + C)/\sqrt{\delta\epsilon})$. $S$ is the set up cost of the quantum process, $U$ is the update cost for one step of the walk and $C$ is the checking cost.*

The set up cost of the Magniez-Nayak algorithm is $2(\ell - 1)$ and update cost is $O(\log k)$. Combined with Szegedy's theorem, some calculation shows that the query cost is minimized at $\ell = k^{2/3}$ and the quantum query complexity is $O(k^{2/3} \log k)$.

## 3.1 Multilinear identity testing (MIT)

Now we are ready to describe our result for multilinear identity testing for a given black-box ring. We start with describing the problem first. Let $R$ be a black-box ring given by a set of additive generators $\{r_1, r_2, \cdots, r_k\}$ and $f(x_1, x_2, \cdots, x_m)$ over $R$ be a multilinear polynomial also given by a black-box. Our problem is to test whether $f(a_1, \cdots, a_m) = 0$ for all $a_i \in R$.

The first step is a suitable generalization of Pak's lemma. For any $i \in [m]$, consider the set $R_i \subseteq R$ defined as follows:

$$R_i = \{u \in R \mid \forall(b_1, \cdots, b_{i-1}, b_{i+1}, \cdots, b_m) \in R^{m-1}, f(b_1, \cdots, b_{i-1}, u, b_{i+1}, \cdots, b_m) = 0\}$$

Clearly, if $f$ is not a zero function from $R^m \to R$, then $|R_i| < |R|$. In the following lemma, we prove that if $f$ is not a zero function then $|R_i| \leq |R|/2$.

**Lemma 3.3** *Let $R$ be any finite ring and $f(x_1, x_2, \cdots, x_m)$ be a multilinear polynomial over $R$ (commuting or noncommuting) such that $f = 0$ is not an identity for $R$. For $i \in [m]$ define*

$$R_i = \{u \in R \mid \forall(b_1, \cdots, b_{i-1}, b_{i+1}, \cdots, b_m) \in R^{m-1}, f(b_1, \cdots, b_{i-1}, u, b_{i+1}, \cdots, b_m) = 0\}. \tag{1}$$

*Then $R_i$ is an additive coset of a proper additive subgroup of $R$ and hence $|R_i| \leq |R|/2$.*

*Proof.* Write $f = A(x_1, \cdots, x_{i-1}, x_i, x_{i+1}, \cdots, x_m) + B(x_1, \cdots, x_{i-1}, x_{i+1}, \cdots, x_m)$ where $A$ is the sum of all the monomials of $f$ containing $x_i$ and $B$ is the sum of the rest of the monomials. Let $v_1, v_2$ be any two distinct elements in $R_i$. Then for any fixed $\bar{y} = (y_1, \cdots, y_{i-1}, y_{i+1}, \cdots, y_m) \in R^{m-1}$, consider the evaluation of $A$ and $B$ over $(y_1, \cdots, y_{i-1}, v_1, y_{i+1}, \cdots, y_m)$ and $(y_1, \cdots, y_{i-1}, v_2, y_{i+1}, \cdots, y_m)$ respectively. For convenience, we abuse the notation and write,

$$A(v_1, \bar{y}) + B(\bar{y}) = A(v_2, \bar{y}) + B(\bar{y}) = 0.$$

$\bar{y}$ is an assignment to $x_1, x_2, \cdots, x_{i-1}, x_{i+1}, \cdots, x_k$ and $v_1, v_2$ are the assignments to $x_i$ respectively. Note that, as $f$ is a multilinear polynomial, the above relation in turns implies that $A(v_1 - v_2, \bar{y}) = 0$.

Consider the set $\hat{R}_i$, defined as follows: fix any $u^{(i)} \in R_i$.

$$\hat{R}_i = \{x - u^{(i)} \mid x \in R_i\}$$

We claim that $\hat{R}_i$ is an (additive) subgroup of $R$. We only need to show that $\hat{R}_i$ is closed under the addition (of $R$). Consider $(x_1 - u^{(i)}), (x_2 - u^{(i)}) \in \hat{R}_i$. Then $(x_1 - u^{(i)}) + (x_2 - u^{(i)}) = (x_1 + x_2 - u^{(i)}) - u^{(i)}$. It is now enough to show that for any $\bar{y} \in R^{m-1}$, $f(x_1 + x_2 - u^{(i)}, \bar{y}) = 0$ ($x_1 + x_2 + u^{(i)}$ is an assignment to $x_i$). Again using the fact that $f$ is multilinear, we can easily see the following:

$$f(x_1 + x_2 - u^{(i)}, \bar{y}) = A(x_1, \bar{y}) + A(x_2, \bar{y}) - A(u^{(i)}, \bar{y}) + B(\bar{y}) = A(x_2, \bar{y}) - A(u^{(i)}, \bar{y}) = A(x_2 - u^{(i)}, \bar{y}) = 0.$$

Note that the last equality follows because $x_2$ and $u$ are in $R_i$. Hence we have proved that $\hat{R}_i$ is a subgroup of $R$. So $R_i = \hat{R}_i + u^{(i)}$ i.e $R_i$ is a coset of $\hat{R}_i$ inside $R$. Also $|R_i| < |R|$ ($f$ is not identically zero over $R$). Thus, finally we get $|R_i| = |\hat{R}_i| \leq |R|/2$. ∎

Our quantum algorithm is based on the algorithm of [MN05]. In the rest of the paper we denote by $S_\ell$ the set of all $\ell$ size subsets of $\{1, 2, \cdots, k\}$ [1]. We follow a quantization of a random walk on $S_\ell \times \cdots \times S_\ell = S_\ell^m$. For $u = \{u_1, u_2, \cdots, u_\ell\}$, define $r_u = r_{u_1} + \cdots + r_{u_\ell}$. Now, we suitably adapt the Lemma 1 of [MN05] in our context. Let $R$ be a finite ring given by a additive generating set $S = \{r_1, \cdots, r_k\}$. W.l.o.g, assume that $r_1$ is the zero element of $R$. Let $\hat{R}$ be a proper additive subgroup of $(R, +)$. Let $j$ be the least integer in $[k]$ such that $r_j \notin \hat{R}$. Since $\hat{R}$ is a proper subgroup of $R$, such a $j$ always exists.

**Lemma 3.4** *Let $\hat{R} < R$ be a proper additive subgroup of $R$ and $T$ be an additive coset of $\hat{R}$ in $R$. Then* $\text{Prob}_{u \in S_\ell}[r_u \notin T] \geq \frac{1-p}{2}$, *where* $p = \frac{\ell(\ell-1) + (k-\ell)(k-\ell-1)}{k(k-1)}$.

*Proof.* Let $j$ be the least integer in $[k]$ such that $r_j \notin \hat{R}$. Since $\hat{R}$ is a proper subgroup of $R$, such a $j$ always exists. Fix a set $u$ of size $\ell$ such that $1 \in u$ and $j \notin u$. Denote by $v$ the set obtained from $u$ by deleting 1 and inserting $j$. This define a one to one correspondence (matching) between all such pair of $(u, v)$. Moreover $r_v = r_u + r_j$ (notice that $r_1 = 0$). Then at least one of the element $r_u$ or $r_v$ is not in $T$. For otherwise $(r_v - r_u) \in \hat{R}$ implying $r_j \in \hat{R}$, which is a contradiction.

Therefore,

$$\text{Prob}_{u \in S_\ell}[r_u \in T \mid j \in u \text{ xor } 1 \in u] \leq \frac{1}{2}.$$

---

[1] Notice that in [MN05], the author consider the set of all $\ell$ tuples instead of subsets. This is important for them as they work in nonabelian structure in general (where order matters). But we will be interested only over additive abelian structure of a ring and thus order does not matter for us.

For any two indices $i, j$,

$$\text{Prob}_{u \in S_\ell}[i, j \in u \text{ or } i, j \notin u] = \frac{\ell(\ell - 1) + (k - \ell)(k - \ell - 1)}{k(k - 1)} = p.$$

Thus,

$$\text{Prob}_{u \in S_\ell}[r_u \in T] \leq (1 - p)/2 + p \leq (1 + p)/2.$$

This completes the proof. ∎

Let $T = R_i$ in Lemma 3.4, where $R_i$ is as defined in Lemma 3.3.

Suppose $f = 0$ is not an identity for the ring $R$. Then, using Lemma 3.4 we show for $u_1, u_2, \cdots, u_m$ picked uniformly at random from $S_\ell$ that $f(r_{u_1}, \cdots, r_{u_m})$ is non zero with non-negligible probability. This is analogous to [MN05, Lemma 2].

**Lemma 3.5** *Let $f(x_1, \cdots, x_m)$ be a multilinear polynomial (in commuting or noncommuting indeterminates) over $R$ such that $f = 0$ is not an identity for the ring $R$. Then,*

$$\text{Prob}_{u_1, \cdots, u_m \in S_\ell}[f(r_{u_1}, \cdots, r_{u_m}) \neq 0] \geq \left(\frac{1 - p}{2}\right)^m.$$

*Proof.* For $i \in [m]$, let $R_i$ be the additive coset defined in Equation 1 of Lemma 3.3. The proof is by simple induction on $m$. The proof for the base case of the induction (i.e for $m = 1$) follows easily from the definition of $R_i$ and Lemma 3.4. By induction hypothesis assume that the result of this lemma holds for all $t$-variate multilinear polynomials $g$ such that $g = 0$ is not an identity for $R$ with $t \leq m - 1$.

Consider the given multilinear polynomial $f(x_1, x_2, \cdots, x_m)$. Then by the Lemma 3.3, $R_m$ is a coset of an additive subgroup $\hat{R}_m$ inside $R$. Pick $u_m \in S_\ell$ uniformly at random. If $f = 0$ is not an identity on $R$ then by Lemma 3.4 we get $r_{u_m} \notin R_m$ with probability at least $\frac{1-p}{2}$. Let $g(x_1, x_2, \cdots, x_{m-1}) = f(x_1, \cdots, x_{m-1}, r_{u_m})$. Since $r_{u_m} \notin R_m$ with probability at least $\frac{1-p}{2}$, it follows that $g = 0$ is not an identity on $R$ with probability at least $\frac{1-p}{2}$. Then, by induction hypothesis, $\text{Prob}_{u_1, \cdots, u_{m-1} \in S_\ell}[g(r_{u_1}, \cdots, r_{u_{m-1}}) \neq 0] \geq \left(\frac{1-p}{2}\right)^{m-1}$. Hence we get, $\text{Prob}_{u_1, \cdots, u_m \in S_\ell}[f(r_{u_1}, \cdots, r_{u_m}) \neq 0] \geq \left(\frac{1-p}{2}\right)^m$, which proves the lemma. ∎

We observe two simple consequences of Lemma 3.5. Notice that $\frac{1-p}{2} = \frac{\ell(k-\ell)}{k(k-1)}$. Letting $\ell = 1$ we get $\frac{1-p}{2} = 1/k$, and Lemma 3.5 implies that if $f = 0$ is not an identity for $R$ then $f(a_1, \cdots, a_m) \neq 0$ for one of the $k^m$ choices for the $a_i$ from the generating set $\{r_1, \cdots, r_k\}$.

**Corollary 3.6** *There is a deterministic $k^m$ query algorithm for* MIT, *where $f$ is $m$-variate and $R$ is given by an additive generating set of size $k$.*

Letting $\ell = k/2$ in Lemma 3.5 we get $\frac{1-p}{2} \geq 1/4$. Hence we obtain the following randomized test which makes $4^m mk$ queries.

**Corollary 3.7** *There is a randomized $4^m mk$ query algorithm for* MIT *with constant success probability, where $f$ is $m$-variate and $R$ is given by an additive generating set of size $k$.*

**Remark.** Corollary 3.6 can be seen as a generalization of the $k^2$ query deterministic test for commutativity. Likewise, Corollary 3.7 is analogous to Pak's $O(k)$ query randomized test for commutativity.

We use Lemma 3.5 to design our quantum algorithm. Our quantum algorithm is based on a quantization of a random walk on $S_\ell^m$ and motivated by the one described in [MN05]. The Lemma 3.5 is used to guarantee that there will at least $\left(\frac{1-p}{2}\right)^m$ fraction of *marked points* in the space $S_\ell^m$ i.e the points where $f$ evaluates to non zero.

Now we describe the random walk on $S_\ell^m$ which is the main building block of our quantum algorithm. In fact we only describe the random walk on $S_\ell$. Over $S_\ell^m$, the random walk consists of just $m$ independent simultaneous random walks on $S_\ell$.

### 3.1.1 Random walk on $S_\ell$

Our random walk can be described as a random walk over a graph $G = (V, E)$ which we define as follows: The vertices of $G$ are all possible $\ell$ subsets of $[k]$. Two vertices are connected by an edge whenever the corresponding sets differ by exactly one element. Notice that $G$ is a connected $\ell(k - \ell)$-regular graph. Also $G$ is well known in the literature as Johnson Graph (with parameter $(k, \ell, \ell - 1)$) [BCN89]. Let $P$ be the normalized adjacency matrix of $G$ with rows and columns are indexed by the subsets of $[k]$. Then $P_{XY} = 1/\ell(k - \ell)$ if $|X \cap Y| = \ell - 1$ and 0 otherwise. It is well known that the spectral gap $\delta$ of $P$ ($\delta = 1 - \lambda$, where $\lambda$ is the second largest eigenvalue of $P$) is $\Omega(1/\ell)$ for $\ell \leq k/2$ [BCN89]. Now we describe the random walk on $G$.

Let the current vertex is $u = \{u_1, u_2, \cdots, u_\ell\}$ and $r_u = r_{u_1} + r_{u_2} + \cdots + r_{u_\ell}$. With probability $1/2$ stay at $u$ and with probability $1/2$ do the following: randomly pick $u_i \in u$ and $j \in [k] \setminus u$. Then move to vertex $v$ such that $v$ is obtained from $u$ by removing $u_i$ and inserting $j$. Compute $r_v$ by simply subtracting $r_{u_i}$ from $r_u$ and adding $r_j$ to it. That will only cost 2 oracle access. Staying in any vertex with probability $1/2$ ensures that the random walk is ergodic. So the stationary distribution of the random walk is always uniform. It is easy to see that the transition matrix of the random walk is $A = (I + P)/2$ where $I$ is the identity matrix of suitable dimension. So the spectral gap of the transition matrix $A$ is $\hat{\delta} = (1 - \lambda)/2 = \delta/2$.

Now, in the following theorem we present the analysis of the query complexity.

**Theorem 3.8** *Let $R$ be a finite ring given as an oracle and $f(x_1, \cdots, x_m)$ be a multilinear polynomial over $R$ given as a black-box. Moreover let $\{r_1, \cdots, r_k\}$ is a given additive generating set for $R$. Then the quantum query complexity of identity testing of $f$ is $O(m(1 + \alpha)^{m/2} k^{\frac{m}{m+1}})$ assuming $k \geq (1 + 1/\alpha)^{m+1}$.*

*Proof.* Our algorithm analysis is similar to the analysis of [MN05].

**Setup cost(S):** For the quantum walk step we need to start with an uniform distribution on $S_\ell^m$. With each $u \in S_\ell$, we maintain a quantum register $|d_u\rangle$ that computes $r_u$. So we need to prepare the following state $|\Psi\rangle$:

$$|\Psi\rangle = \frac{1}{\sqrt{|S_\ell^m|}} \sum_{u_1, u_2, \cdots, u_m \in S_\ell^m} |u_1, r_{u_1}\rangle \otimes |u_2, r_{u_2}\rangle \otimes \cdots \otimes |u_m, r_{u_m}\rangle.$$

It is easy to see that to compute any $r_{u_j}$ we need $\ell - 1$ oracle access to the ring oracle. Since in each of $m$ independent walk, quantum queries over all choices of $u$ will be made in parallel (using quantum superposition), the total query cost for setup is $m(\ell - 1)$.

**Update cost(U):** It is clear from the random walk described in the section 3.1.1, that the update cost over $S_\ell$ is only 2 oracle access. Thus for the random walk on $S_\ell^m$ which is just $m$ independent random walks, one on each copy of $S_\ell$, we need a total update cost $2m$.[2]

---

[2]In [MN05] the underlying group operation is not necessarily commutative (it is being tested for commutativity). Thus the update cost is more.

**Checking cost(C):** To check whether $f$ is zero on a point during the walk, we simply query the oracle for $f$ once.

Recall from Szegedy's result [Sze04] (as stated in Theorem 3.2), the total cost for query complexity is $Q = S + \frac{1}{\sqrt{\hat{\delta}\epsilon}}(U + C)$ where $\epsilon = \left(\frac{1-p}{2}\right)^m$ is the proportion of the marked elements and $\hat{\delta}$ is the spectral gap of the transition matrix $A$ described in section 3.1.1. Combining together we get, $Q \leq m\left[(\ell - 1) + \frac{3}{\sqrt{\hat{\delta}\epsilon}}\right]$. From the random walk described in the section 3.1.1, we know that $\hat{\delta} \geq \frac{1}{2\ell}$. Hence, $Q \leq m\left[(\ell - 1) + \frac{3\sqrt{2\ell}}{\left(\frac{1-p}{2}\right)^{\frac{m}{2}}}\right]$. Notice that, $\frac{1-p}{2} = \frac{\ell}{k}\left(\frac{1-\frac{\ell}{k}}{1-\frac{1}{k}}\right)$. Substituting for $\frac{1-p}{2}$ we get, $Q \leq m\left[(\ell - 1) + 3\sqrt{2}k^{m/2}\frac{1}{\ell^{\frac{m-1}{2}}\left(\frac{k-\ell}{k-1}\right)^{m/2}}\right]$. We will choose a suitably small $\alpha > 0$ so that $\frac{k-1}{k-\ell} < 1 + \alpha$. Then we can upper bound $Q$ as follows. $Q \leq m\left[(\ell - 1) + 3\sqrt{2} \cdot (1+\alpha)^{m/2}k^{m/2}\frac{1}{\ell^{\frac{m-1}{2}}}\right]$. Now our goal is to minimize $Q$ with respect to $\ell$ and $\alpha$. For that we choose $\ell = k^t$ where we will fix $t$ appropriately in the analysis. Substituting $\ell = k^t$ we get, $Q \leq m\left[(k^t - 1) + 3\sqrt{2} \cdot (1+\alpha)^{m/2}t^{1/2}k^{\frac{m-(m-1)t}{2}}\right]$. Choosing $t = (m/(m+1))$, we can easily see that the query complexity of the algorithm is $O(m(1+\alpha)^{m/2}k^{\frac{m}{m+1}})$. Finally, recall that we need choose an $\alpha > 0$ so that $\frac{k-1}{k-\ell} \leq 1 + \alpha$. Clearly, it suffices to choose $\alpha$ so that $(1+\alpha)\ell \leq \alpha k$. Letting $\ell = k^{m/m+1}$ we get the constraint $(1 + 1/\alpha)^{m+1} \leq k$ which is satisfied if $e^{(m+1)/\alpha} \leq k$. We can choose $\alpha = \frac{m+1}{\ln k}$. ∎

**Remark.** The choice of $\alpha$ in the above theorem shows some trade-offs in the query complexity between the parameters $k$ and $m$. For constant $m$ notice that this gives us an $O(k^{m/m+1})$ query upper bound for the quantum algorithm.

Finally, it is easy to observe that the quantum algorithm and its analysis given in Theorem 3.8 hold for a more general problem stated in the following theorem.

**Theorem 3.9** *Let $R$ be a black-box finite ring given by ring oracle and suppose $A = \langle r_1, r_2, \cdots, r_k \rangle$ is an additive subgroup of $R$ given by generators $r_i \in R$. Let $f(x_1, x_2, \cdots, x_m)$ be a black-box multilinear polynomial $f : R^m \to R$. There is a quantum algorithm with query complexity $O(m(1+\alpha)^{m/2}k^{\frac{m}{m+1}})$ (assuming $k \geq (1+1/\alpha)^{m+1}$), to check if $f = 0$ is an identity for the additive abelian group $A$.[3]*

# 4   A reduction from m-SPLIT COLLISION problem

We can easily show that any classical algorithm for the MIT problem must make $\Omega(k)$ queries. This is an easy consequence of observations in [MN05]. Specifically, an $\Omega(k)$ is shown in [MN05] for commutativity testing of a black-box group $G$ given by $k$ generators. It is a consequence of the randomized query complexity for the UNIQUE SPLIT COLLISION problem. The lower bound argument applies to MIT as well, implying an $\Omega(k)$ query lower bound for the problem.

We do not have an explicit lower bound result for the quantum query complexity of multilinear identity testing problem (MIT) on rings. However, in this section we show that the more general problem of multilinear identity testing for additive subgroups (described in Theorem 3.9) is at least as hard as m-SPLIT COLLISION, which is a version of the m-COLLISION problem.

---

[3]I.e. checking if $f(a_1, \cdots, a_m) = 0$ for all $a_i \in A$.

We first describe some automata theory that is useful for our reduction.

## 4.1 Automata theory background

We recall some standard automata theory notation (see, for example, [HU78]). Fix a finite automaton $A = (Q, \Sigma, \delta, q_0, q_f)$ which takes as input strings in $\Sigma^*$. $Q$ is the set of states of $A$, $\Sigma$ is the alphabet, $\delta : Q \times \Sigma \to Q$ is the transition function, and $q_0$ and $q_f$ are the initial and final states respectively (throughout, we only consider automata with unique accepting states). For each letter $b \in \Sigma$, let $\delta_b : Q \to Q$ be the function defined by: $\delta_b(q) = \delta(q, b)$. These functions generate a submonoid of the monoid of all functions from $Q$ to $Q$. This is the transition monoid of the automaton $A$ and is well-studied in automata theory: for example, see [Str94, page 55]. We now define the 0-1 matrix $M_b \in \mathbb{F}^{|Q| \times |Q|}$ as follows: $M_b(q, q') = 1$ if $\delta_b(q) = q'$, and 0 otherwise.

The matrix $M_b$ is simply the adjacency matrix of the graph of the function $\delta_b$. As the entries of $M_b$ are only zeros and ones, we can consider $M_b$ to be a matrix over any field $\mathbb{F}$.

Furthermore, for any $w = w_1 w_2 \cdots w_k \in \Sigma^*$ we define the matrix $M_w$ to be the matrix product $M_{w_1} M_{w_2} \cdots M_{w_k}$. If $w$ is the empty string, define $M_w$ to be the identity matrix of dimension $|Q| \times |Q|$. For a string $w$, let $\delta_w$ denote the natural extension of the transition function to $w$; if $w$ is the empty string, $\delta_w$ is simply the identity function. It is easy to check that: $M_w(q, q') = 1$ if $\delta_w(q) = q'$ and 0 otherwise. Thus, $M_w$ is also a matrix of zeros and ones for any string $w$. Also, $M_w(q_0, q_f) = 1$ if and only if $w$ is accepted by the automaton $A$. We now describe the reduction.

**Theorem 4.1** *The* m-SPLIT COLLISION *problem reduces to multilinear polynomial identity testing (*MIT*) for additive subgroups of black-box rings.*

*Proof.* An instance of m-SPLIT COLLISION is a function $f : [k] \to [k]$ given as an oracle, where we assume w.l.o.g. that $k = nm$. Divide $\{1, 2, \cdots, k\}$ into $m$ intervals $I_1, I_2, \cdots, I_m$, each containing $n$ consecutive points of $[k]$. Recall that $f$ has an $m$-collision if for some $j \in [k]$ we have $|f^{-1}(j)| = m$. Furthermore, $f$ is said to have an $m$-split collision if for some $j \in [k]$ we have $|f^{-1}(j)| = m$ and $|f^{-1}(j) \cap I_i| = 1$ for each interval $I_i$.

Consider the alphabet $\Sigma = \{b, c, b_1, b_2, \cdots, b_m\}$. For each $i \in [k]$, define the $k$-tuple $r_i$ over $\Sigma$ as follows: $r_i[i] = b$ and $r_i[f(i)] = b_j$ where $i \in I_j$. For an index $s \in [k] \setminus \{i, f(i)\}$ define $r_i[s] = c$.

Let $\mathcal{A} = (Q, \Sigma, \delta, q_0, q_f)$ be a deterministic finite state automaton that accepts all strings $w \in \Sigma^*$ such that each $b_j, 1 \leq j \leq m$ occurs at least once in $w$. It is easy to see that such an automaton with a single final state $q_f$ can be designed with total number of states $|Q| = 2^{O(m)} = t$. W.l.o.g. let the set of states $Q$ be renamed as $\{1, 2, \cdots, t\}$, where 1 is the initial state and $t$ is the final state.

For each letter $a \in \Sigma$, let $M_a$ denote the $t \times t$ transition matrix for $\delta_a$ (as defined in Section 4.1). Since each $M_a$ is a $t \times t$ 0-1 matrix, each $M_a$ is in the ring $\mathcal{M}_t(\mathbb{F}_2)$ of $t \times t$ matrices with entries from the field $\mathbb{F}_2$. Let $R$ denote the $k$-fold product ring $(\mathcal{M}_t(\mathbb{F}_2))^k$. Clearly, $R$ is a finite ring (which is going to play the role of the black-box ring in our reduction). We now define an additive subgroup $T$ of $R$, where we describe the generating set of $T$ using the m-SPLIT COLLISION instance $f$.

For each index $i \in [k]$, define an $k$-tuple $T_i \in R$ as follows. Let $T_i[i] = M_b$, $T_i[f(i)] = M_{b_j}$ (where $i \in I_j$) and for each index $s \notin \{i, f(i)\}$ define $T_i[s] = M_c$. The additive subgroup of $R$ we consider is $T = \langle T_1, T_2, \cdots, T_k \rangle$ generated by the $T_i, 1 \leq i \leq k$.

Furthermore, define two $t \times t$ matrices $A$ and $B$ in $\mathcal{M}_t(\mathbb{F}_2)$ as follows. Let $A[1, 1] = 1$ and $A[u, \ell] = 0$ for $(u, \ell) \neq (1, 1)$. For the matrix $B$, let $B[t, 1] = 1$ and $B[u, \ell] = 0$ for $(u, \ell) \neq (t, 1)$.

**Claim 4.2** *Let $w = w_1 w_2 \cdots w_t \in \Sigma^*$ be any string. Then the automaton $\mathcal{A}$ defined above accepts $w$ if and only if the matrix $A M_{w_1} M_{w_2} \cdots M_{w_t} B$ is nonzero.*

*Proof of Claim* By definition of the matrices $M_a$, the $(1, k)^{th}$ entry of the product $M_{w_1} M_{w_2} \cdots M_{w_t}$ is 1 if and only if $w$ is accepted by $\mathcal{A}$. By definition of the matrices $A$ and $B$ the claim follows immediately.

Now, consider the polynomial $P(x_1, x_2, \cdots, x_m)$ in noncommuting indeterminates $x_1, \cdots, x_m$ with coefficients from the matrix ring $R$ defined as follows:

$$P(x_1, x_2, \cdots, x_m) = \bar{A} x_1 x_2 \cdots x_m \bar{B},$$

where $\bar{A} = (A, A, \ldots, A) \in R$ and $\bar{B} = (B, B, \cdots, B) \in R$ are $k$-tuples of $A$'s and $B$'s respectively. We claim that the multilinear polynomial $P(x_1, x_2, \cdots, x_m) = 0$ is an identity for the additive subgroup $T$ if and only if $f$ has no $m$-split collision.

**Claim 4.3** $P(x_1, \cdots, x_m) = 0$ *is an identity for the ring $T = \langle T_1, \cdots, T_k \rangle$ if and only if $f$ has no $m$-split collision.*

*Proof of Claim* Suppose $f$ has an $m$-split collision. Specifically, let $i_j \in I_j$ $1 \leq j \leq m$ be indices such that $f(i_1) = \cdots = f(i_m) = \ell$. In the polynomial $P$, we substitute for indeterminate $x_j$ by $T_{i_j}$ for $1 \leq j \leq m$. Consider the product $M = T_{i_1} \cdots T_{i_m}$ in the ring $T$. This product is an $k$-tuple of $t \times t$ matrices such that in the $\ell^{th}$ component $M$ has the matrix $\prod_{t=1}^m M_{b_t}$ where $i_t \in I_t$. Since $b_{i_1} b_{i_2} \cdots b_{i_m} \in \Sigma^*$ is a length $m$-string containing all the $b_j$'s it will be accepted by the automaton $\mathcal{A}$. Consequently, the $(q_0, q_f)^{th}$ entry of the matrix $M$, which is the $(1, k)^{th}$ entry, is 1 (as explained in Section 4.1). It follows that the $(1, 1)$ entry of the matrix $AMB$ is 1. Hence $P = 0$ is not an identity over the ring $T$.

For the other direction, assume that $f$ has no $m$-split collision. We need to show that $P = 0$ is an identity for the ring $T$. For any $m$ elements $S_1, S_2, \cdots, S_m \in T$ consider $P(S_1, S_2, \cdots, S_m) = \bar{A} S_1 S_2 \cdots S_m \bar{B}$. Since Each $S_j$ is an $\mathbb{F}_2$-linear combination of the generators $T_1, \cdots, T_k$, it follows by distributivity in the ring $R$ that $P(S_1, S_2, \cdots, S_m)$ is an $\mathbb{F}_2$-linear combination of terms of the form $P(T_{k_1}, T_{k_2}, \cdots, T_{k_m})$ for some $m$ indices $k_1, \cdots, k_m \in [k]$. Thus, it suffices to show that $P(T_{k_1}, T_{k_2}, \cdots, T_{k_m}) = 0$.

Let $\hat{T} = T_{k_1} T_{k_2} \cdots T_{k_m}$. Then, for each $j \in [k]$ we have $\hat{T}[j] = T_{k_1}[j] T_{k_2}[j] \cdots T_{k_m}[j]$. Since $f$ has no $m$-split collision, for each $j \in [N]$ the set of matrices $\{M_{b_1}, M_{b_2}, \cdots, M_{b_m}\}$ is *not* contained in the set $\{T_1[j], T_2[j], \cdots, T_k[j]\}$. Thus, $\hat{T}[j] = T_{k_1}[j] T_{k_2}[j] \cdots T_{k_m}[j]$ is a product of matrices $M_{w_1} M_{w_2} \cdots M_{w_m}$ for a word $w = w_1 w_2 \cdots w_m$ that is not accepted by $\mathcal{A}$. It follows from the previous claim that $A\hat{T}[j]B = 0$. Hence $P(T_{k_1}, T_{k_2}, \cdots, T_{k_m}) = 0$ which completes the proof. ∎

**Acknowledgment.** We thank Ashwin Nayak for comments and suggestions.